\documentclass{edp-conf}
\usepackage{graphicx}
\begin{document}

\TitreGlobal{SF2A 2001}

\title{ISO's view of SS\,433 Jet Interactions with the Interstellar Medium} 
\runningtitle{ISO's view of W50}
\author{Fuchs Ya\"el}\address{Service d'Astrophysique, CEA/Saclay, 91191 Gif sur Yvette Cedex}
%
\maketitle
\begin{abstract} 
SS\,433 is a jet emitting X-ray binary surrounded by the W50 radio nebula. The
SS\,433\,/\,W50 system is an excellent laboratory for studying relativistic
jet interaction with the surrounding interstellar medium. In this
context, part of the W50 nebula has been mapped with ISOCAM at
15\,$\mu$m. I will show the results particularly on the W50 west lobe,
and on 2 emitting zones detected with IRAS who have also been observed
in millimetre wavelength (CO(1-0) transition), and for one of them by
spectroscopy with ISOLWS and ISOSWS between 2 and 200 $\mu$m.
 \end{abstract}
%
\setcounter{footnote}{1}
\section{Introduction}

  SS\,433 is an X-ray binary probably composed of a high-mass star and
  a neutron star, with a 13 days binary period. The system emits
  relativistic (0.26\,c) jets showing a 162.5 days precession observed
  at subarcsecond scale in radio till \mbox{$\sim 10^{17}$\,cm} from
  SS\,433. At large scale these jets are observed in X-ray begining at
  $\sim 15'$ (14\,pc) from SS\,433, and they are responsible for the
  unusual elongated shape of W50 ($\sim 1^\circ \times 2^\circ$), the
  possible supernova remnant radio nebula around SS\,433. We mapped at
  15\,$\mu$m with \mbox{ISOCAM}\footnote{The ISOCAM data presented in
  this paper was analysed using "CIA", a joint development by the ESA
  Astrophysics Division and the ISOCAM Consortium. The ISOCAM
  Consortium is led by the ISOCAM PI, C. Cesarsky.}, the infrared
  camera on board of the Infrared Space Observatory (ISO), a small
  part of the eastern lobe where an X-ray knot lies, the north-east
  quarter of the central circular part of W50, and nearly all the
  western lobe.
 
\section{Eastern lobe observations}

   There is no particular 15\,$\mu$m emission in the observed east
   parts of W50, and no correlation was found between the IR images
   and the corresponding ones in X-ray (with ROSAT and ASCA) and radio (at
   20\,cm), both for the central field and for the X-ray knot
   area. This is not surprising as the central part of W50 must have
   been swept away from its material by the supernova explosion, and
   the eastern lobe is described as faint in radio and less dense than
   the western one by Dubner et al. (1998). The X-ray knot, the
   brightest knot seen by ROSAT at 0.1--2.4\,keV and called ``e2'' by
   Safi-Harb and \"Ogelman (1997), is coincident with an optical
   filament and is probably due to shocks with the supernova remnant
   (Safi-Harb and Petre 1999), but it is not visible at 15\,$\mu$m
   with ISOCAM sensitivity.

\section{Western lobe observations}

   The western lobe map at 14--16\,$\mu$m (see Fig.\,1) shows two
   main emitting regions aligned with the relativistic jet direction
   predicted by the kinematic model (Margon 1984 and references there
   in) based on the radio compact jets observations, and seen at large
   scale in X-ray (see Fig.\,2). The western X-ray lobe was observed with ROSAT
   (0.1--2.4\,keV) by Brinkmann et al. (1996), and with ROSAT and ASCA
   (0.5--9\,keV) by Safi-Harb and \"Ogelman (1997) who could not
   conclude if its emission is thermal or not. It is only partly
   coincident with the nearest ISOCAM emission region from
   SS\,433 (see Fig.\,2). Its soft X-ray emission fades and its hard X-ray one
   disappears where the second emitting region begins.  This farest IR
   emitting region from SS\,433 coincides with the radio
   ``ear'' emission at the western edge of W50.\\
\begin{figure}[!ht]
\includegraphics[width=12cm]{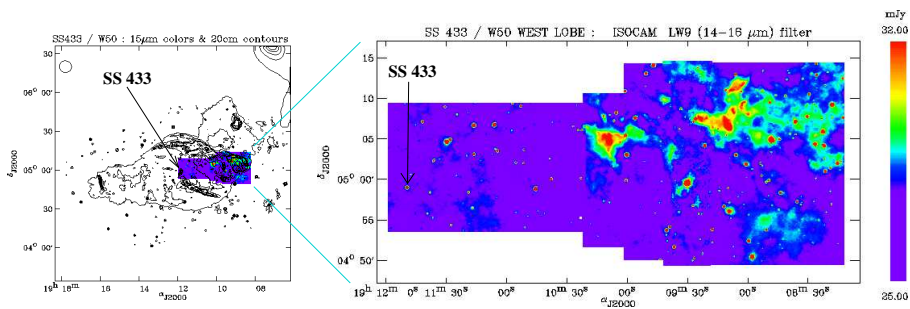}
\caption{Left: location of the ISOCAM map in the W50 western lobe; W50 is shown at 20\,cm in contours. Right: ISOCAM 14--16\,$\mu$m image with a $-\sigma$ to +2\,$\sigma$ scale around the median flux value, and with a $6'' \times 6''$ per pixel field of view. SS\,433 is the point source at the left edge of this image as indicated.}
\label{fig1}
\end{figure}
\begin{figure}[!ht]
\includegraphics[width=12cm]{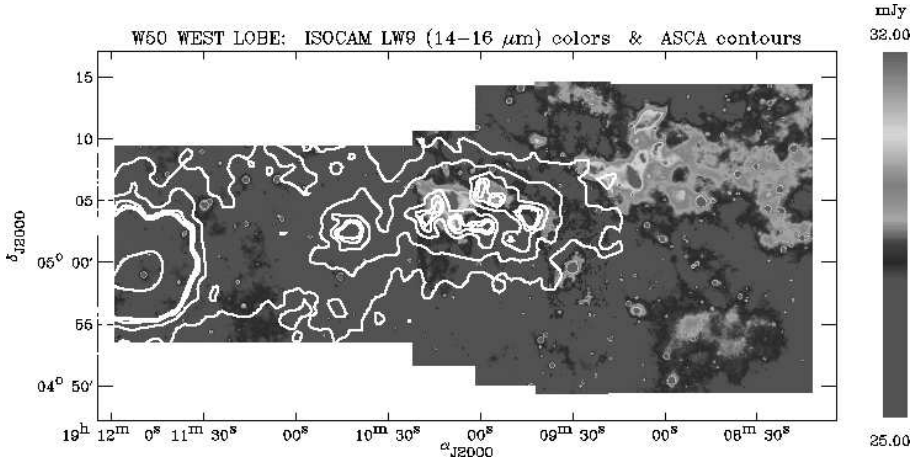}
\caption{Superimposition of the western lobe X-ray ASCA emission in contours with the ISOCAM 15\,$\mu$m image, showing that the IR emissions lies along the same axis as the X-ray lobe.}
\label{fig.2}
\end{figure}

   These two ISOCAM emitting regions correspond to IR knots observed
   with IRAS at 12, 25, 60 and 100\,$\mu$m by Band (1987), and named
   ``knot 2'' for the nearest region from SS\,433 and ``knot 3'' for
   the farest (see Fig.\,3). The $6'' \times 6''$ per pixel ISOCAM spatial
   resolution enables to reveal their structure for the first
   time. The 15\,$\mu$m knots emission is diffuse with punctual or not
   ``hot spots''.\\

   No emission from the CO(1-0) transition line at 115\,GHz is
   observed between SS\,433 and ``knot 2'', as in IR at 15\,$\mu$m in
   the ISOCAM image. Two \mbox{CO(1-0)} emitting regions are
   coincident with the two IRAS knots as shown in Fig.\,3 (data kindly
   provided by Durouchoux, private communication). Their Doppler
   velocity is $\sim 50$\,km.s$^{-1}$ which corresponds to the same
   distance as W50. Their shapes are similar to the ISOCAM shapes, so the
   IR and millimetre emissions are unlikely coincident by chance but
   must be physically linked. So the observed ISOCAM emissions are due
   to regions lying inside W50. \\
\begin{figure}[!ht]
\includegraphics[width=12cm]{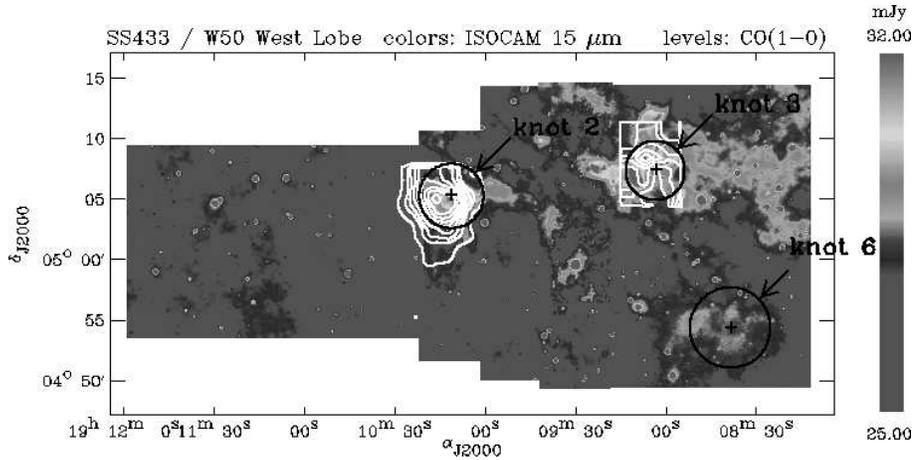}
\caption{Superimposition of the CO(1--0) emitting line contours at 115\,GHz (2.6\,mm) with ISOCAM 15\,$\mu$m image. The IRAS knots locations are indicated. ``knot 2'' and ``knot 3'' are emitting regions in both IR and millimetre with similar shapes. ``knot 6'' lies just at the border of W50 and was not observed at millimetre wavelengths.}
\label{fig.3}
\end{figure}

   ``Knot 3'' was observed with ISOSWS and ISOLWS, respectively the
   Short and Long-Wavelength Spectrometers on board of the ISO
   satellite. SWS observation consists in several small
   wavelength-range spectra between 2 and 40 $\mu$m, which do not
   reveal any emission lines when the flux is not under the detection
   limit. The LWS continuum spectrum, between 40 and 200 $\mu$m, is
   consistent above 60\,$\mu$m with optically thin or optically thick
   thermal emission from dust at $\sim 30$\,K. Below 60\,$\mu$m, the
   spectrum is flatter due to an additionnal component thermal or not.\\
	
   Interpretation of the two knots emission is uncertain due to the
   lack of observations in near and mid-infrared wavelengths where it could be
   thermal or not. The oberved IR emission could be due to dust
   regions heated by young stars still embedded in their molecular
   birth clouds. These regions could be crossed by the relativistic
   western jet, which then could heat the dust with collisions between
   its energetic particles and the denser material of these
   clouds. Otherwise, the mid-infrared emission could be of
   synchrotron nature, coming from the very energetic particles of the
   jet, reaccelerated by shocks with the western lobe denser
   medium. These particles would lose their energy by synchrotron
   radiation in X-ray, then in IR as they travel further from SS\,433.\\

   More multi-wavelength observations, are planed\,: in near-infrared
   in order to reveal shocked regions or young star regions, in
   millimetre in order to map the whole western lobe and to know the
   molecular clouds limits, in X-ray with XMM-Newton to distinguish
   between thermal and non thermal emission and thanks to the better
   spatial resolution to compare precisely the X-ray images with the
   other wavelengths ones. These new data will help to better
   understand the emission mecanisms from these large scale
   relativistic jets, and they could also help understanding
   extragalactic jets from AGN as there are larger scales replica.



\end{document}